\documentclass[usenatbib]{mnras}
\usepackage{graphicx}
\usepackage{amsmath}
\usepackage{amssymb}
\usepackage{color}
\usepackage{url}

\usepackage{threeparttable}

\newcommand\lsim{\mathrel{\rlap{\lower4pt\hbox{\hskip1pt$\sim$}}
        \raise1pt\hbox{$<$}}}
\newcommand\gsim{\mathrel{\rlap{\lower4pt\hbox{\hskip1pt$\sim$}}
        \raise1pt\hbox{$>$}}}
\newcommand{\lya}{\ifmmode\mathrm{Ly}\alpha\else{}Ly$\alpha$\fi}
\newcommand{\lyb}{\ifmmode\mathrm{Ly}\beta\else{}Ly$\beta$\fi}
\newcommand{\igm}{\ifmmode\mathrm{IGM}\else{}IGM\fi}
\newcommand{\lae}{\ifmmode\mathrm{LAE}\else{}LAE\fi}
\newcommand{\h}{\ifmmode\mathrm{H}\else{}H\fi}
\newcommand{\hi}{\ifmmode\mathrm{H\,{\scriptscriptstyle I}}\else{}H\,{\scriptsize I}\fi}
\newcommand{\hii}{\ifmmode\mathrm{H\,{\scriptscriptstyle II}}\else{}H\,{\scriptsize II}\fi}
\newcommand{\cmb}{\ifmmode\mathrm{CMB}\else{}CMB\fi}
\newcommand{\qso}{\ifmmode\mathrm{QSO}\else{}QSO\fi}
\newcommand{\eor}{\ifmmode\mathrm{EoR}\else{}EoR\fi}
\newcommand{\heii}{\ifmmode\mathrm{He\,{\scriptscriptstyle II}}\else{}He\,{\scriptsize II}\fi}
\newcommand{\heiii}{\ifmmode\mathrm{He\,{\scriptscriptstyle III}}\else{}He\,{\scriptsize III}\fi}
\newcommand{\ciii}{\ifmmode\mathrm{C\,{\scriptscriptstyle III]}}\else{}C\,{\scriptsize III]}\fi}
\newcommand{\oiii}{\ifmmode\mathrm{O\,{\scriptscriptstyle III}}\else{}O\,{\scriptsize III}\fi}
\newcommand{\aliii}{\ifmmode\mathrm{Al\,{\scriptscriptstyle III}}\else{}Al\,{\scriptsize III}\fi}
\newcommand{\mgii}{\ifmmode\mathrm{Mg\,{\scriptscriptstyle II}}\else{}Mg\,{\scriptsize II}\fi}
\newcommand{\fe}{\ifmmode\mathrm{Fe}\else{}Fe\fi}
\newcommand{\nv}{\ifmmode\mathrm{N\,{\scriptscriptstyle V}}\else{}N\,{\scriptsize V}\fi}
\newcommand{\niv}{\ifmmode\mathrm{N\,{\scriptscriptstyle IV]}}\else{}N\,{\scriptsize IV]}\fi}
\newcommand{\cii}{\ifmmode\mathrm{C\,{\scriptscriptstyle II}}\else{}C\,{\scriptsize II}\fi}
\newcommand{\civ}{\ifmmode\mathrm{C\,{\scriptscriptstyle IV}}\else{}C\,{\scriptsize IV}\fi}
\newcommand{\siv}{\ifmmode\mathrm{Si\,{\scriptscriptstyle IV}}\else{}Si\,{\scriptsize IV}\fi}
\newcommand{\siii}{\ifmmode\mathrm{Si\,{\scriptscriptstyle II}}\else{}Si\,{\scriptsize II}\fi}
\newcommand{\siiii}{\ifmmode\mathrm{Si\,{\scriptscriptstyle III]}}\else{}Si\,{\scriptsize III]}\fi}
\newcommand{\ovi}{\ifmmode\mathrm{O\,{\scriptscriptstyle VI}}\else{}O\,{\scriptsize VI}\fi}
\newcommand{\sioiv}{\ifmmode\mathrm{Si\,{\scriptscriptstyle IV}\,\plus O\,{\scriptscriptstyle IV]}}\else{}Si\,{\scriptsize IV}\,+O\,{\scriptsize IV]}\fi}

\newcommand{\cmfst}{\textsc{\small 21CMFAST}}

\pdfoutput=1
\title[Large-volume reionisation simulations]{Generating extremely large-volume reionisation simulations}
\author[B. Greig et al.] {Bradley~Greig$^{1,2}$\thanks{E-mail:~greigb@unimelb.edu.au}, J. Stuart B. Wyithe$^{1,2}$, Steven G. Murray$^{3}$, Simon J. Mutch$^{1,2}$\newauthor \& Cathryn M. Trott$^{2,4}$ \\
$^1$School of Physics, University of Melbourne, Parkville, VIC 3010, Australia \\
$^2$ARC Centre of Excellence for All-Sky Astrophysics in 3 Dimensions (ASTRO 3D) \\
$^3$School of Earth and Space Exploration, Arizona State University, Tempe, AZ \\
$^4$International Centre for Radio Astronomy Research (ICRAR), Curtin University, Bentley, WA, Australia \\
}

\begin{document}
\maketitle \begin{abstract}
\noindent
Preparing for the first detection of the cosmic 21-cm signal from large-scale interferometer experiments requires rigorous testing of the data analysis and reduction pipelines. To validate that these pipelines do not erroneously remove or add features that can mimic the cosmic signal (e.g. from side-lobes or large-scale power leakage), we require reionisation simulations larger than the experiments primary field of view. For an experiment such as the MWA, with a field of view of $\sim25^{2}$~deg.$^{2}$, this would require a simulation of several Gpcs, which is currently infeasible. To overcome this, we developed a simplified version of the semi-numerical reionisation simulation code \cmfst{} preferencing large volumes over some physical accuracy by assuming linear theory for structure formation. With this, we constructed a 7.5 Gpc comoving volume with voxel resolution of $\sim1.17$ cMpc tailored specifically to the binned spectral resolution of the MWA. This simulation was used for validating the pipelines for the 2020 MWA 21-cm power spectrum (PS) upper limits (Trott et al.). We then use this large-volume simulation to explore: (i) whether smaller volume simulations are biased by the missing large-scale modes, (ii) non-Gaussianity in estimates of the cosmic variance, (iii) biases in the recovered 21-cm PS following foreground wedge removal and (iv) the impact of tiling smaller volume simulations to achieve extremely large volumes. In summary, we find: (i) no biases from missing large-scale power, (ii) significant contribution from non-Gaussianity in the cosmic variance as expected following Mondal et al. (iii) an over-estimate of the 21-cm PS of 10-20 per cent following wedge mode excision for our particular model and (iv) tiling smaller volume simulations under-estimates the large-scale power and also the estimated cosmic variance.
\end{abstract} 

\section{Introduction}

Directly observing the first stars and primordial galaxies is rendered nigh on impossible by the inescapable neutral hydrogen fog that enshrouds the early Universe. This fog is gradually lifted in the intergalactic medium (IGM) as the neutral hydrogen is ionised by the cumulative output of ultra-violet (UV) photons from stars and galaxies. This phase transition is referred to as the Epoch of Reionisation (EoR).

Our most promising method to observe the EoR is through detecting the 21-cm hyperfine spin-flip transition of neutral hydrogen. This faint signal, measured relative to the Cosmic Microwave Background (CMB), is in either emission or absorption depending on both the ionisation and thermal state of the IGM \citep[see e.g.][]{Gnedin:1997p4494,Madau:1997p4479,Shaver:1999p4549,Tozzi:2000p4510,Gnedin:2004p4481,Furlanetto:2006p209,Morales:2010p1274,Pritchard:2012p2958}. Crucially, by measuring the 21-cm signal we gain a 2D picture of the spatial distribution of the neutral hydrogen in the IGM. Being a line transition, repeating this over a broad frequency range builds up full 3D time-dependent movie of IGM in the early Universe. From this, we can infer the properties of the galaxies responsible for driving the EoR.

To detect this 3D signal we require large-scale interferometer experiments, which are specifically designed to be sensitive to the spatial fluctuations. The first generation of these experiments include the Low-Frequency Array (LOFAR; \citealt{vanHaarlem:2013p200}), the Murchison Wide Field Array (MWA; \citealt{Tingay:2013p2997,Wayth:2018}), the Precision Array for Probing the Epoch of Reionisation (PAPER; \citealt{Parsons:2010p3000}), the Owens Valley Radio Observatory Long Wavelength Array (OVRO-LWA; \citealt{Eastwood:2019}) and the upgraded Giant Metrewave Radio Telescope (uGMRT; \citealt{Gupta:2017}). Next generation experiments, offering larger collecting areas and lower noise, are currently underway or are planned such as the Hydrogen Epoch of Reionization Array (HERA; \citealt{DeBoer:2017p6740}), NenuFAR (New extension in Nan\c{c}ay Upgrading loFAR; \citealt{Zarka:2012}) and the Square Kilometre Array (SKA; \citealt{Mellema:2013p2975,Koopmans:2015}).

Although these interferometer experiments are yet to detect the 21-cm signal\footnote{A detection of excess absorption has been reported by the Experiment to Detect the Global EoR Signature (EDGES; \citealt{Bowman:2018}) global signal experiment, however, its cosmological origins have been disputed in the literature \citep[see e.g.][]{Hills:2018,Draine:2018,Bowman:2018b,Bradley:2019,Singh:2019,Singh:2022}}, recent improvements in flagging of cleaned data and the data analysis and reduction pipelines have yielded the strongest yet upper-limits on the 21-cm power spectrum (PS) from LOFAR \citep{Mertens:2020}, the MWA \citep{Trott:2020} and HERA \citep{HERA:2022a}.

Crucially, 21-cm signal detection is a precision science. The signal is extremely faint, hidden behind astrophysical foregrounds roughly five orders of magnitude brighter than the cosmic signal. Thus, when developing data analysis and reduction pipelines it is vital to test all components that could erroneously inject or remove signal. One vital piece towards this is having sufficiently large simulation volumes that extend well beyond the primary beam and into the side lobes. For example, for an experiment such as the MWA, with a field-of-view of $\sim25^{2}$~deg.$^{2}$ at 150 MHz ($z\sim8.5$), this corresponds to a transverse size of $\sim4$~Gpc. Simulations of this scale are currently infeasible.

In this work, we introduce a modified version of the semi-numerical simulation code \cmfst{} \citep{Mesinger:2007p122,Mesinger:2011p1123} specifically tailored towards generating sufficiently large reionisation volumes. To achieve this, we make a few simplifying assumptions such as adopting linear structure formation and that the IGM spin temperature is in excess of the CMB temperature ($T_{S} \gg T_{\rm CMB}$). Further, we follow a similar approach to \citet{Greig:2011} and only simulate a restricted volume via considering a frequency dependent depth.

With these large volume simulations, redshift-space distortions, signal evolution and sky curvature effects can be correctly simulated to provide realistic 21-cm datasets to apply to data analysis pipelines. The simulations introduced in this work have been added to the software package, \textsc{WODEN} \citep{Line:2022}, which produce realistic end-to-end simulations for the MWA Telescope, where they provide a reference dataset for testing power spectrum amplitude and slope, as well as a comparison against which different analysis techniques can be made (e.g., if the data cleaning and treatment removes sufficient systematic power to reach the cosmological signal).

With such a large simulation volume we can also test and verify several assumptions and approximations that are typically made when being computationally limited to smaller simulation volumes. For example, the impact on the 21-cm PS when tiling simulations to achieve sufficiently large volumes or whether the 21-cm PS is impacted by large-scale modes longer than the simulation volume. Further, we statistically explore the 21-cm PS by breaking up our large volume simulation into smaller sub-volumes. We explore the impact of the non-Gaussianity in the 21-cm signal on the PS cosmic variance uncertainty following \citet{Mondal:2015,Mondal:2016} and the potential bias in the 21-cm PS when avoiding the contaminated foreground `wedge' modes relative to the true 21-cm PS from the full simulation as highlighted by \citet{Jensen:2016}. 

The outline of this paper is as follows. In Section~\ref{sec:Method} we detail our modified version of \cmfst{} and describe the adopted ionising source prescription. Next, in Section~\ref{sec:Results} we outline the specifically tailored large-volume simulation for the MWA before investigating various properties and assumptions related to the 21-cm PS. Finally, in Section~\ref{sec:Conclusion} we conclude with our closing remarks. Unless stated otherwise, quoted quantities are in co-moving units and we adopt the cosmological parameters:  ($\Omega_\Lambda$, $\Omega_{\rm M}$, $\Omega_b$, $n$, $\sigma_8$, $H_0$) = (0.69, 0.31, 0.048, 0.97, 0.81, 68 km s$^{-1}$ Mpc$^{-1}$), consistent with recent results from the Planck mission \citep{Planck:2020}.

\section{Method} \label{sec:Method}

In this work we use a heavily modified version of the semi-numerical simulation code \cmfst{}\footnote{In particular, the older C-only version, https://github.com/andreimesinger/21cmFAST} \citep{Mesinger:2007p122,Mesinger:2011p1123}. \cmfst{} employs approximate but efficient methods to describe the astrophysics of the EoR which compare favourably to computationally expensive radiative transfer (RT) simulations on scales ($\gtrsim1$ cMpc) relevant to 21-cm interferometer experiments \citep{Zahn:2011p1171}. This efficiency is achieved by computing the ionisation field using the excursion-set approach \citep{Furlanetto:2004p123} which compares the time-integrated number of ionising photons (following a source prescription) to the number of baryons within spherical regions of decreasing radius, $R$. Under this method, a simulation voxel at the co-ordinates ($\boldsymbol{x}, z$) is considered ionised if,
\begin{eqnarray} \label{eq:ioncrit}
n_{\rm ion}(\boldsymbol{x}, z | R, \delta_{R}) \geq 1.
\end{eqnarray}
Here, $n_{\rm ion}$ is the cumulative number of IGM ionising photons per baryon inside a spherical region of size, $R$ and corresponding smoothed overdensity, $\delta_{R}$. Explicitly, in this work we ignore the contributions from inhomogenous recombinations \citep[e.g.][]{Sobacchi:2014p1157} and partial ionisations by X-rays to boost the overall computational efficiency.

For our ionising sources, we adopt the \citet{Park:2019} astrophysical parameterisation for high-$z$ galaxies which directly connects the star-formation rate and ionising escape fraction to the host dark matter halo mass. The advantage of such an approach is that following some simple conversions, UV luminosity functions (LFs) can be produced and compared against high-$z$ observations. For specific details we defer the interested reader to the aforementioned work and provide a brief summary of the parameterisation below. 

First, it is assumed that the typical stellar mass of a galaxy, $M_{\ast}$, can be related to its host halo mass via, 
\begin{eqnarray} \label{}
M_{\ast}(M_{\rm h}) = f_{\ast}\left(\frac{\Omega_{\rm b}}{\Omega_{\rm m}}\right)M_{\rm h},
\end{eqnarray}
and that $f_{\ast}$, the fraction of galactic gas in stars, can be expressed as a power-law in halo mass with index $\alpha_{\ast}$ and normalised at a dark matter halo of mass $10^{10}$~$M_{\odot}$ through $f_{\ast, 10}$, 
\begin{eqnarray} \label{}
f_{\ast} = f_{\ast, 10}\left(\frac{M_{\rm h}}{10^{10}\,M_{\odot}}\right)^{\alpha_{\ast}}.
\end{eqnarray}

Following this, we obtain an estimate for the star-formation rate (SFR) by dividing the stellar mass by a characteristic time-scale, $t_{\ast}$, corresponding to a fraction of the Hubble time, $H^{-1}(z)$,
\begin{eqnarray} \label{}
\dot{M}_{\ast}(M_{\rm h},z) = \frac{M_{\ast}}{t_{\ast}H^{-1}(z)}.
\end{eqnarray}

The escape fraction of UV ionising photons, $f_{\rm esc}$, is also assumed to be related to the halo mass through a power-law of index, $\alpha_{\rm esc}$, and is again normalised at $10^{10}$~$M_{\odot}$ via $f_{\rm esc, 10}$,
\begin{eqnarray} \label{}
f_{\rm esc} = f_{\rm esc, 10}\left(\frac{M_{\rm h}}{10^{10}\,M_{\odot}}\right)^{\alpha_{\rm esc}}.
\end{eqnarray}

Finally, to account for feedback and inefficient cooling processes which limit small mass haloes from hosting active, star-forming galaxies a duty-cycle, $f_{\rm duty}$, is included which describes the fraction, $(1 - f_{\rm duty})$, of dark matter haloes that cannot host star-forming galaxies,
\begin{eqnarray} \label{}
f_{\rm duty} = {\rm exp}\left(-\frac{M_{\rm turn}}{M_{\rm h}}\right),
\end{eqnarray}
with $M_{\rm turn}$ corresponding to this suppression scale. In summary, this ionising source prescription results in six free parameters:  $f_{\ast, 10}$, $\alpha_{\ast}$, $f_{\rm esc, 10}$, $\alpha_{\rm esc}$, $t_{\ast}$ and $M_{\rm turn}$.

The number of ionising photons, $n_{\rm ion}$ inside a spherical region of size, $R$ is then obtained via:
\begin{eqnarray} \label{eq:ioncrit2}
n_{\rm ion} = \bar{\rho}^{-1}_b\int^{\infty}_{0}{\rm d}M_{\rm h} \frac{{\rm d}n(M_{h}, z | R, \delta_{R})}{{\rm d}M_{\rm h}}f_{\rm duty} \dot{M}_{\ast}f_{\rm esc}N_{\gamma/b},
\end{eqnarray}
where $\bar{\rho}_b$ is the mean baryon density, $\frac{{\rm d}n}{{\rm d}M_{\rm h}}$ is the conditional halo mass function (HMF) and $N_{\gamma/b}$ is the number of ionising photons per stellar baryon. Here, we use the Press-Schecter HMF \citep{Lacey:1993p115} normalised to the mean of the Sheth-Tormen HMF \citep{Sheth:1999p2053}, and $N_{\gamma/b}=5000$ representative of a Salpeter initial mass function \citep{Salpeter:1955}. 

The cosmic 21-cm signal is computed via its brightness temperature contrast against the Cosmic Microwave Background (CMB) temperature, $T_{\rm CMB}$ \citep[e.g.][]{Furlanetto:2006p209}:
\begin{eqnarray}
\delta T_{\rm b}(\nu) &=& \frac{T_{\rm S} - T_{\rm CMB}}{1+z}\left(1-e^{-\tau_{\nu_0}}\right) \nonumber \\
&\approx& 27x_{\hi{}}(1+\delta_{\rm nl})\left(\frac{H}{{\rm d}v_{\rm r}/{\rm d}r+H}\right)
\left(1 - \frac{T_{\rm CMB}}{T_{\rm S}}\right) \nonumber \\
& & \times \left(\frac{1+z}{10}\frac{0.15}{\Omega_{\rm m}h^{2}}\right)^{1/2}
\left(\frac{\Omega_{b}h^{2}}{0.023}\right)~{\rm mK},
\end{eqnarray}
where $\tau_{\nu_{0}}$ is the optical depth of the 21-cm line, $\nu_0$, $T_{\rm S}$ is the gas spin temperature, $\delta_{\rm nl}(\boldsymbol{x},z)$ is the evolved (Eularian) overdensity, $x_{\hi{}}$ is the ionisation fraction, $H(z)$ is the Hubble parameter, 
${\rm d}v_{\rm r}/{\rm d}r$ is the gradient of the LOS component of the velocity and all quantities are evaluated at 
redshift $z = \nu_{0}/\nu - 1$. To aid computational efficiency, in this work we ignore the effects of peculiar velocities (except in Section~\ref{sec:wedge}) while also assuming that the gas spin temperature is in excess of the CMB temperature ($T_{S} \gg T_{\rm CMB}$). Thus, we compute the brightness temperature contrast as,
\begin{eqnarray} \label{eq:Tb}
\delta T_{\rm b}(\nu) \approx 27x_{\hi{}}(1+\delta_{\rm nl})\left(\frac{1+z}{10}\frac{0.15}{\Omega_{\rm m}h^{2}}\right)^{1/2}
\left(\frac{\Omega_{b}h^{2}}{0.023}\right)~{\rm mK}.
\end{eqnarray}

Typically, \cmfst{} first generates a high-resolution 3D realisation of the linear density before applying second-order Lagrange perturbation theory \citep[e.g][]{Scoccimarro:1998p7939} to obtain the requisite velocity and evolved density fields. It then smooths the evolved fields onto a lower resolution grid to mitigate numerical artefacts. However, to achieve extremely large-volume reionisation simulations, we bypass this step and consider only the linear density field. 

Even following this, our largest achievable simulation size is limited to the maximum available memory on our computer network. This limitation is driven by the necessity of performing 3D Fast-Fourier Transforms (FFTs). To bypass this, we follow the approach of \citet{Greig:2011} and take a considerable hit to computational efficiency by storing the 3D data on hard-disk and performing successive 1D and 2D FFTs rather than the expensive 3D FFTs in memory. In effect, this limits our memory requirements to $N^{2}$ rather than $N^{3}$ and takes advantage of the fact that hard-disk space exceeds total memory (since it is cheaper, though considerably slower to access). Importantly, by adopting linear theory (initial conditions are generated in Fourier space) and highlighting that the excursion-set algorithm is applied in Fourier space, we can also minimise the number of FFTs between real and Fourier space. Finally, at no point do we ever generate a full 3D real-space cubic volume, rather we instead truncate the depth to a desired frequency bandwidth. That is, we consider a much smaller line-of-sight depth to our volumes than the transverse direction. By doing this, we only need to perform a limited number of 1D and 2D FFTs to achieve our requisite volume. However, at all times the statistics of the field remain correct as we always fully sample the largest scale modes.

Importantly, although we have made several assumptions to boost computational efficiency at the expense of some physical accuracy, our overall goal is to have a method to generate extremely large 21-cm simulations for testing data analysis and reduction pipelines. For these, physical accuracy is not necessarily the most important requirement, instead, we simply require that the statistics and correlations in our large volume simulations be representative of those expected of the true signal over an extremely large range of spatial scales and survey footprint. Additionally, for these extremely large volume simulations we do not include the effects of sky curvature at present. In future we will look into the impact of sky-curvature on large-volume simulations.

\section{Results} \label{sec:Results}

\subsection{Tailored MWA simulation}

\begin{figure*} 
	\begin{center}
	  \includegraphics[trim = 0.2cm 0.6cm 0cm 0.5cm, scale = 0.5]{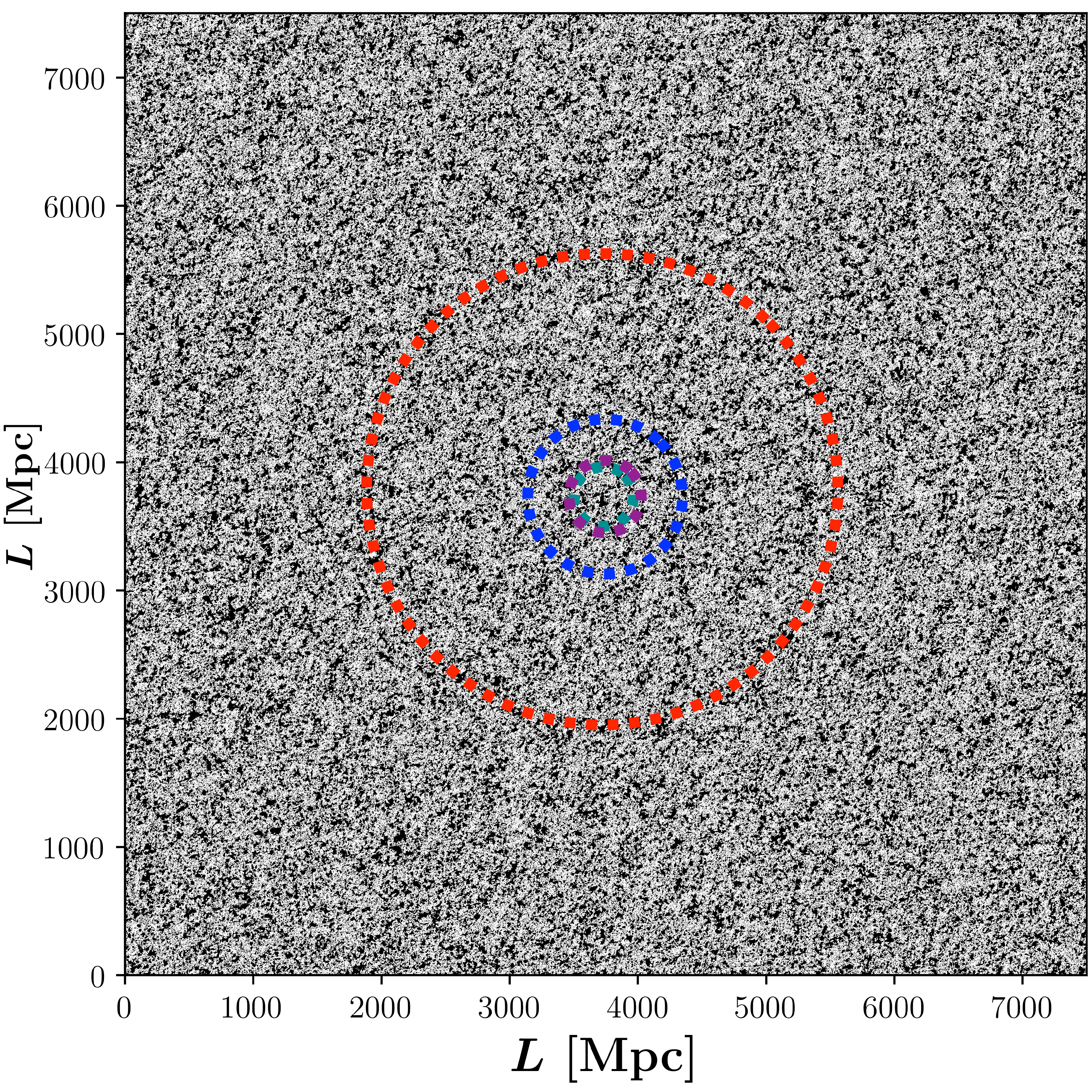}
	\end{center}
\caption[]{A 2D slice of the ionisation field, $x_{\hi{}}$, obtained from the 7.5 comoving Gpc, 6400 voxel simulation at $z=8$ (cell resolution $\sim1.17$~cMpc). The coloured dashed circles correspond to the field-of-view for a variety of 21-cm interferometers: MWA (red), HERA (blue), LOFAR (purple) and SKA (teal). See Table~\ref{tab:FoV} for further details.}
\label{fig:LargeSlice}
\end{figure*}

The \citet{Trott:2020} MWA upper-limits were determined from observations spanning 137--197 MHz in two slightly overlapping 30.72 MHz bands. Over these bands, the MWA frequency resolution is $\sim80$~kHz, which sets the minimum spatial resolution for our large-volume simulations and the maximal line-of-sight length required (384 channels). At 150~MHz ($z\sim8.5$), the MWA field of view is $\sim25^{2}$~${\rm deg.}^{2}$, corresponding to a comoving side-length of $\sim4$~Gpc. To exceed these requirements, we generate a single simulation volume with a comoving transverse side-length of 7.5~Gpc over 6400 voxels and 600 voxels ($703.1$~Mpc) along the line-of-sight direction\footnote{Such a simulation required $\sim6$ days, using 32 CPUs and $\sim10$~GB of memory. The majority of this time is spent reading/writing to file, which could be considerably improved using solid state hard-drives.}. This simulation corresponds to a cell resolution of $\sim1.17$~cMpc. To our knowledge, this is the largest volume 21-cm simulation in the literature\footnote{As a verification of our method we also performed a simulation with a transverse side-length of 9.6 Gpc and 8092 voxels.}. Using this simulation, we construct two 21-cm light-cones spanning the observing bands ($z=6.2 - 7.5$ and $z=7.5 - 9.4$) by stitching together 2D slices of the 21-cm signal from our simulation volume by linearly interpolating in cosmic time (e,g, \citealt{Datta:2012p7679,Datta:2014p4990,LaPlante:2014p7651,Ghara:2015p7650,Mondal:2018}).

For this simulation, we adopt $f_{\ast, 10} = 0.05$, $\alpha_{\ast} = 0.5$, $f_{\rm esc, 10} = 0.08$, $\alpha_{\rm esc} = -0.5$, $t_{\ast} = 0.5$ and $M_{\rm turn} = 10^{8.7}$, consistent with the fiducial model adopted in \citet{Park:2019}, which was shown to match all recent observational constraints on the reionisation epoch. Additionally, in neglecting inhomogeneous recombinations we are required to set an effective mean free path for the ionising photons inside the ionised regions, $R_{\rm mfp}$. We adopt $R_{\rm mfp} = 15$~comoving Mpc motivated by the fiducial model in \citet{Greig:2015p3675}.

In Figure~\ref{fig:LargeSlice} we present a 2D slice of the IGM ionisation fraction, $x_{\hi{}}$, at $z=8$ for which we obtain an IGM neutral fraction of $\bar{x}_{\hi{}}\sim0.55$. As a visual demonstration of the physical extent of these simulations, we overlay the respective fields of view for each of the MWA (red), HERA (blue), LOFAR (purple) and SKA (teal). Note that unlike the MWA, LOFAR and SKA which use phase tracking to follow a single patch of the sky, HERA is a drift-scan experiment (sky rotates over the fixed zenith pointing dishes), thus, here we show its instantaneous field of view. In Table~\ref{tab:FoV} we provide the collecting area of a single station/element and the corresponding field of view for each 21-cm interferometer experiment.

\begin{table*}
%\tiny
\begin{tabular}{@{}lcccc}
\hline
Instrument & Collecting area (m$^{2}$) & Field of View (deg.$^2$) & Comoving distance\\
 & (single element) & (@150 MHz) & (Mpc) \\
\hline
\vspace{0.8mm}
MWA  & 21.5 & 24.7$^{2}$ & 3977.3\\
\hline
\vspace{0.8mm}
LOFAR & 745 & 3.7$^{2}$ & 598.8 \\
\hline
\vspace{0.8mm}
HERA  & 154 & 8.2$^{2}$ & 1317.3\\
\hline
\vspace{0.8mm}
SKA  & 1165 & 3.0$^2$ & 479.0\\
\hline
\end{tabular}
\caption{A summary of the collecting area of a single element/station for each of the 21-cm interferometer experiments, the corresponding field of view at 150~MHz and the transverse comoving distance of the simulation required to match the field of view at 150 MHz. Note for HERA, we show the instantaneous field of view as it operates in a drift-scan mode.}
\label{tab:FoV}
\end{table*}

\subsection{Exploring large-scale modes} \label{sec:lsm}

Typically, simulations tailored for a 21-cm interferometer experiment will be designed to match the physical extent of the instruments primary beam. Although problematic for full tests of end-to-end pipelines, for most usage cases this assumption should be sufficient. However, by only simulating a volume equivalent to the primary beam, these simulations forgo modes on much longer scales which would otherwise be present in the observed 21-cm signal. In principle, the absence of these large-scale modes could result in an under-estimate of the true large-scale power. 

We can test the impact of neglecting modes beyond the physical extent of the primary beam by using our 7.5 Gpc simulation as a reference simulation to represent the true Universe. We then compare the 3D spherically averaged 21-cm power spectrum (PS) obtained from our 7.5 Gpc simulation to the PS obtained from the smaller, independent realisations of the 21-cm signal tailored to match each interferometer experiment. Throughout, we define the 21-cm PS as $\Delta^{2}_{21}(k,z) \equiv k^{3}/(2\pi^{2}V)\,\delta \bar{T}^{2}_{\rm b}(z)\,\langle |\delta_{21}(\boldsymbol{k},z)|^{2}\rangle_{k}$
 where $\delta_{21}(\boldsymbol{x},z) \equiv \delta T_{\rm b}(\boldsymbol{x},z)/\bar{\delta T_{\rm b}}(z) -1$. Further, when calculating the 21-cm PS we neglect all $k_{\perp} = 0$ modes. That is, we remove all pure line-of-sight modes as these are not visible to an interferometer\footnote{Note, this is somewhat of a crude approximation as interferometers do in fact have a response to these $k_{\perp} = 0$ modes, although it is suppressed due to the tapered response. While this does not impact the results of this work, realistically one should include the true instrument response rather than taking this approximation when comparing simulations to observations.} \citep[e.g.][]{Datta:2012p7679}.

\begin{figure*} 
	\begin{center}
	  \includegraphics[trim = 0.2cm 0.6cm 0cm 0.3cm, scale = 0.95]{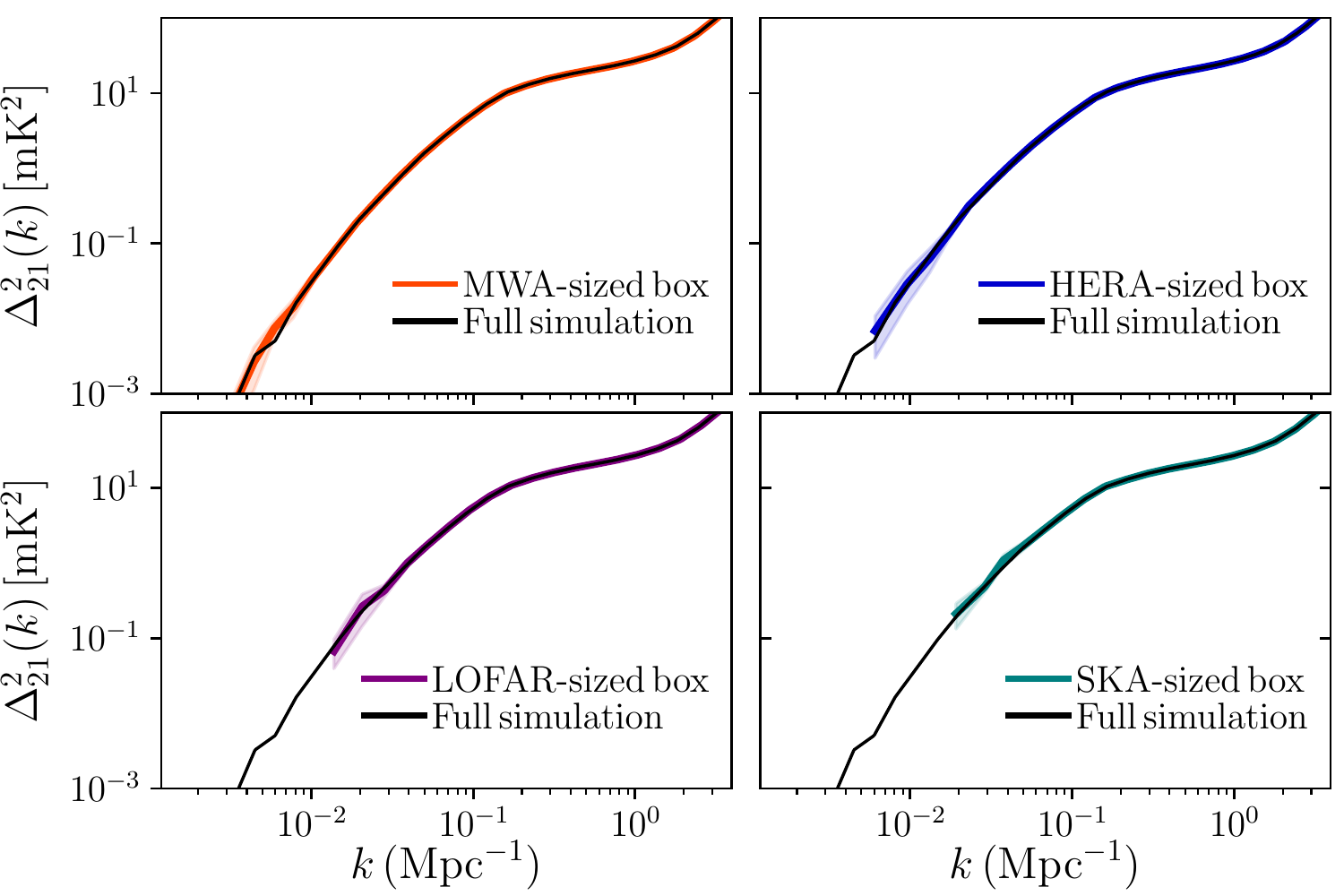}
	\end{center}
\caption[]{The 21-cm power spectrum from simulations tailored to the field-of-view of specific 21-cm interferometers. The black curve corresponds to the 21-cm power spectrum from the full $7500 \times 7500 \times 500$ Mpc simulation. The red, blue, purple and teal curves correspond to simulation volumes with angular extents equivalent to the field-of-view of the MWA, LOFAR, HERA and the SKA, respectively. The shaded region corresponds to the 1$\sigma$ scatter from 10 realisations with different initial conditions.}
\label{fig:LargeModes}
\end{figure*}

In Figure~\ref{fig:LargeModes} we compare the 21-cm PS obtained from our 7.5 Gpc simulation (black curve) to the mean and 68th percentile scatter obtained from 10 independent realisations of the 21-cm signal. In each panel we compare the 21-cm PS obtained from the primary field of view of the MWA (red), HERA (blue), LOFAR (purple) and SKA (teal). In all cases, we compute the 21-cm PS over a simulation volume with the same observing bandwidth of $\sim30$MHz (transverse distance is equivalent to the field of view).

Immediately from Figure~\ref{fig:LargeModes} we see no obvious deviations at large scales. The mean 21-cm PS determined over the 10 independent realisations for each survey footprint overlaps with the 21-cm PS from the 7.5 Gpc simulation. On the largest scales for each individual instrument, we observe some scatter in the 21-cm PS owing to cosmic variance, where the binned 21-cm power has few modes (i.e. Poisson sampling error). Thus, provided the simulation volume exceeds the field of view, this should not be a concern. We can then conclude that omitting large-scales modes beyond the primary field of view does not alter the recovered statistics. This is consistent with the hybrid, large-volume ($>1$~Gpc) reionisation simulations generated by \citet{Kim:2016}, whereby the amplitude of the large-scale 21-cm PS converges for increasing simulation volumes beyond 500~Mpc.

Note, this investigation slightly differs from previous works by \citet{Iliev:2014}, \citet{LaPlante:2014p7651} and \citet{Kaur:2020}. Here, in each case the authors explore sub-volumes from their largest reionisation simulation, which in the case of \citet{Iliev:2014} and \citet{Kaur:2020} is used to determine a minimum volume for reionisation simulations to achieve convergence of the large-scale radiative effects. Since this is a sub-division of a larger volume, these smaller simulations 
contain modes beyond the scale of the sub-volume. In our work, we generate new independent realisations of our smaller volume simulations that do not include modes beyond the scale of the simulation. 

\subsection{Exploring cosmic variance} \label{sec:CV}

The 3D spherically averaged 21-cm PS is calculated by averaging over all Fourier modes that fall within spherical shells. Naturally, the variance of this statistic is then given by the Poisson sampling error based on the total number of Fourier modes in each spherical shell,
\begin{eqnarray} \label{eq:CV}
\sigma_{\Delta^{2}_{21}}(k, z) = \Delta^{2}_{21}(k, z)\sqrt{\frac{(2\pi)^{2}}{V k^{2}\Delta k}}.
\end{eqnarray}
Here, $\Delta^{2}_{21}(k, z)$ is the dimensional 21-cm PS measured within a volume, $V$, with spherical shells separated by $\Delta k$.

However, this assumes that the 21-cm PS fully encodes all the information about the 21-cm signal (i.e. that the 21-cm signal is Gaussian). Instead it is well known that the 21-cm signal is highly non-Gaussian \citep[e.g.][]{Furlanetto:2004p123,Furlanetto:2004b,Morales:2004,Cooray:2005}, in which case the higher-order moments of the 21-cm signal will be non-zero. As a result of this \citet{Mondal:2016} have shown that the cosmic variance uncertainty on the 21-cm PS is instead,
\begin{eqnarray} \label{eq:CVwT}
\sigma_{\Delta^{2}_{21}}(k, z) = \Delta^{2}_{21}(k, z)\sqrt{\frac{(2\pi)^{2}}{V k^{2}\Delta k}} + \sqrt{\frac{T(k, k)}{V}},
\end{eqnarray}
where $T(k, k)$ is the Trispectrum component which arises from the four-point correlation function of the brightness temperature fluctuations. 

To explore this non-Gaussianity in the 21-cm PS cosmic variance, we extract a large number of sub-volumes from our large 7.5 Gpc simulation. In particular, we extract sub-volumes equivalent in spatial extent to the field-of-view for each 21-cm interferometer experiment (as listed in Table~\ref{tab:FoV}), each with the same 30 MHz bandwidth along the line-of-sight. To create these sub-volumes, we first split the full $7.5\times7.5$~Gpc$^{2}$ sky-area into our requisite smaller volumes without overlapping regions. Then, to increase our number statistics, we generate an inset sky-area offset from the simulation edge by half the size of the respective instrument field-of-view. We then additionally split this inset area into non-overlapping sub-volumes. In doing so, we obtain a total of 5, 41, 221 and 365 different sub-volumes from our single, large volume simulation for the MWA, HERA, LOFAR and the SKA, respectively. By considering sub-volumes equivalent to each instruments field-of-view, we consider volumes $\sim2-10\times$ larger than those considered by \citet{Mondal:2016} and \citet{Mondal:2017}. Further, by splitting up our large volume simulation we can obtain up to $\sim7$ times as many realisations, increasing our statistical sampling of the numerical cosmic variance uncertainty, albeit at the expense of notably less physics rich simulations.

\begin{figure*} 
	\begin{center}
	  \includegraphics[trim = 0.2cm 0.6cm 0cm 0.5cm, scale = 1.]{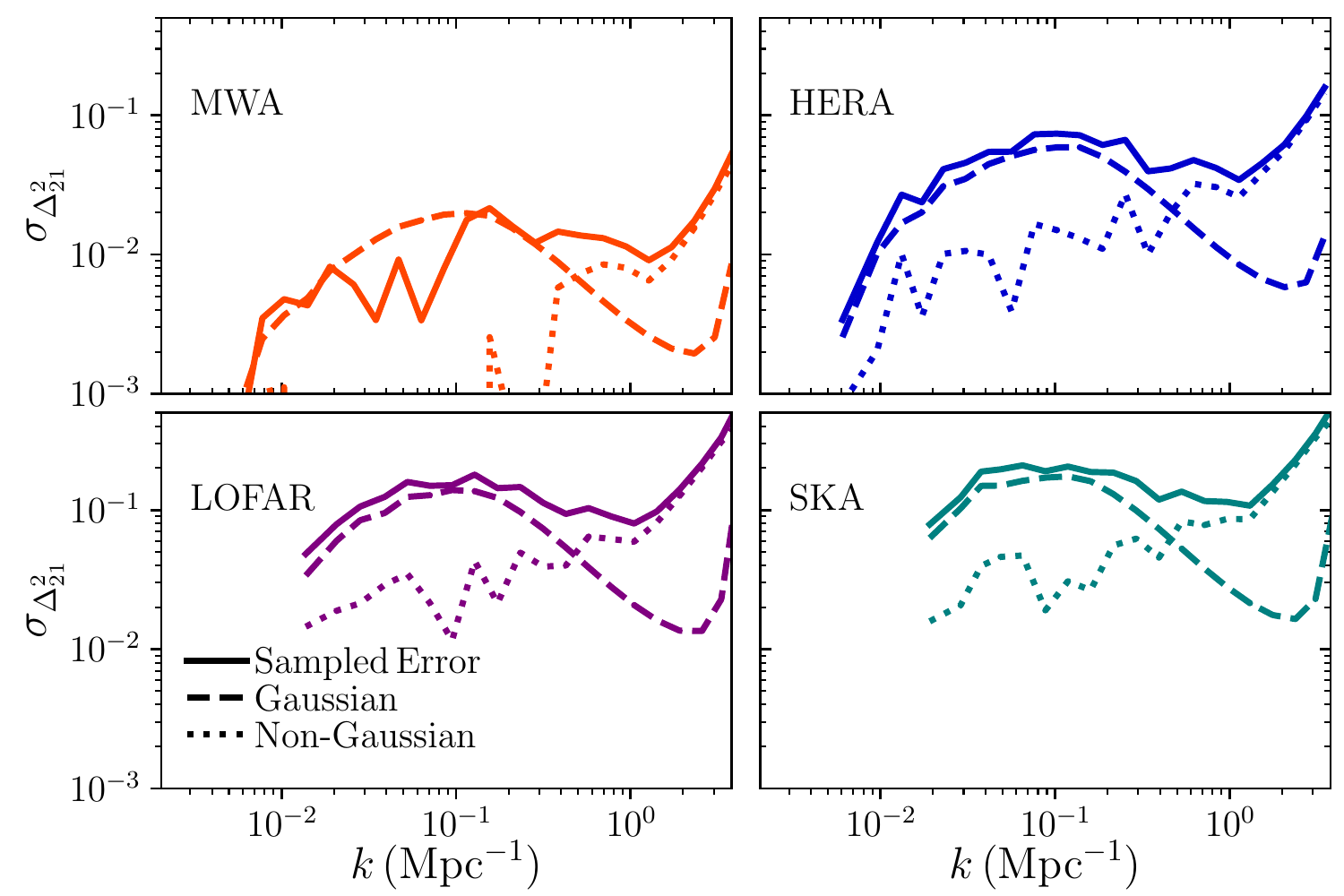}
	\end{center}
\caption[]{The $1\sigma$ variation in the 21-cm power spectrum estimated from sub-volumes of size equivalent to the 21-cm experiment field-of-view (solid curves) compared to the theoretically expected variation assuming Gaussian errors (dashed) and the non-Gaussian contribution to the sample variance (dotted). \textit{Top left:} MWA, \textit{Top right:} HERA, \textit{Bottom left:} LOFAR and \textit{Bottom right:} SKA.}
\label{fig:Variation}
\end{figure*}

In Figure~\ref{fig:Variation} we show the resultant $1\sigma$ uncertainty on the 21-cm PS obtained from our distribution of sub-volumes (solid curves) for each 21-cm interferometer experiment. In contrast, we also show the expected theoretical estimate of the uncertainty from Equation~\ref{eq:CV} if we assumed Gaussian statistics (dashed curves). The dotted curve corresponds to the non-Gaussian (Trispectrum) component of the cosmic-variance as highlighted in Equation~\ref{eq:CVwT}.

On relatively large scales ($k<0.5$~Mpc$^{-1}$), we find that the true cosmic variance uncertainty is well approximated by a Gaussian component for the theoretically expected error. However, progressing to intermediate and smaller scales, the amplitude of the non-Gaussianity increases significantly, with the Trispectrum component dominating by $\geq10\times$ that of the Gaussian term. This transition scale of $k>0.5$~Mpc$^{-1}$ for a non-Gaussian dominated cosmic variance uncertainty on the 21-cm PS is consistent with that of \citet{Mondal:2016} and \citet{Mondal:2017} for a similar stage of reionisation ($\bar{x}_{\hi{}}\sim0.55$). On these scales, as the EoR progresses, the growth and percolation of the ionised regions drives the non-Gaussianity in the cosmic 21-cm signal. Importantly, we find the relative amplitude (the amplitude simply scales as $\propto V^{-1}$) and shape of the full cosmic variance uncertainty to be consistent across the differently sized sub-volumes. This indicates no additional sources of cosmic variance uncertainty due to the finite-size of the simulation volume considered.

Note however, that while the non-Gaussian term for the 21-cm PS cosmic variance begins to dominate on scales larger than $k>0.5$~Mpc$^{-1}$, this need not be too concerning for 21-cm experiments. Typically, on these scales, the 21-cm PS will be dominated by the instrument thermal noise (see figure 2 of \citealt{Greig:2020b}) making these two terms comparable. Nevertheless, care must be taken when estimating the 21-cm PS cosmic variance error.

\subsection{Impact of foreground `wedge' avoidance on cosmic variance} \label{sec:wedge}

The cosmic 21-cm signal observed by radio interferometer experiments is measured in visibility space ($uv$ coverage), which is the 2D Fourier transform of the brightness signal in the image (sky) plane. Measuring the power spectrum then requires gridding these visibilities. However, these visibilities are frequency dependent. Line-of-sight power in Fourier space (frequency dependent) can then leak into the transverse (frequency independent) Fourier modes. This results in the well known `wedge' feature in cylindrical 2D Fourier space, where measured power within this regime is contaminated \citep{Datta:2010p2792,Vedantham:2012p2801,Morales:2012p2828,Parsons:2012p2833,Trott:2012p2834,Thyagarajan:2013p2851,Liu:2014p3465,Liu:2014p3466,Thyagarajan:2015p7294,Thyagarajan:2015p7298,Pober:2016p7301,Murray:2018}. While it is plausible to mitigate or `clean' these contaminated modes (see e.g.\ \citealt{Chapman:2019} for a comprehensive review), it is typically easier to simply avoid this contaminated region entirely, measuring the power spectrum in the relatively pristine region above the `wedge'.

The spatial location of this `wedge' in 2D cylindrical space is determined by,
\begin{eqnarray} \label{eq:wedge}
k_{\parallel} =  mk_{\perp} + b
\end{eqnarray} 
where $k_{\parallel}$ and $k_{\perp}$ are the line-of-sight and transverse Fourier modes, respectively. The constant $b$ corresponds to an additive buffer region above the `wedge' of $\Delta k_{\parallel} = 0.1 \,h$~Mpc$^{-1}$ extending beyond the horizon limit while,
\begin{eqnarray}
m = \frac{D_{\rm C}H_{0}E(z){\rm sin}(\theta)}{c(1+z)}.
\end{eqnarray} 
Here, $D_{\rm C}$ is the comoving distance, $H_{0}$ is the Hubble constant, $E(z) = \sqrt{\Omega_{\rm m}(1+z)^{3} + \Omega_{\Lambda}}$ and $\theta$ is the angular radius of the field of view which we conservatively take as $\theta = \pi/2$ (i.e. observing down to the horizon).

Previously, we numerically estimated the cosmic variance in the 21-cm PS measured over the full 30 MHz bandwidth simulation volume for each interferometer experiment. To highlight the impact of `wedge' avoidance on estimating the 21-cm PS, in particular on the amplitude of the statistical uncertainty, we perform the same analysis as in Section~\ref{sec:CV} instead calculating the 21-cm PS using only Fourier modes above the `wedge'. 

\begin{figure*} 
	\begin{center}
	  \includegraphics[trim = 0.2cm 0.6cm 0cm 0.5cm, scale = 0.95]{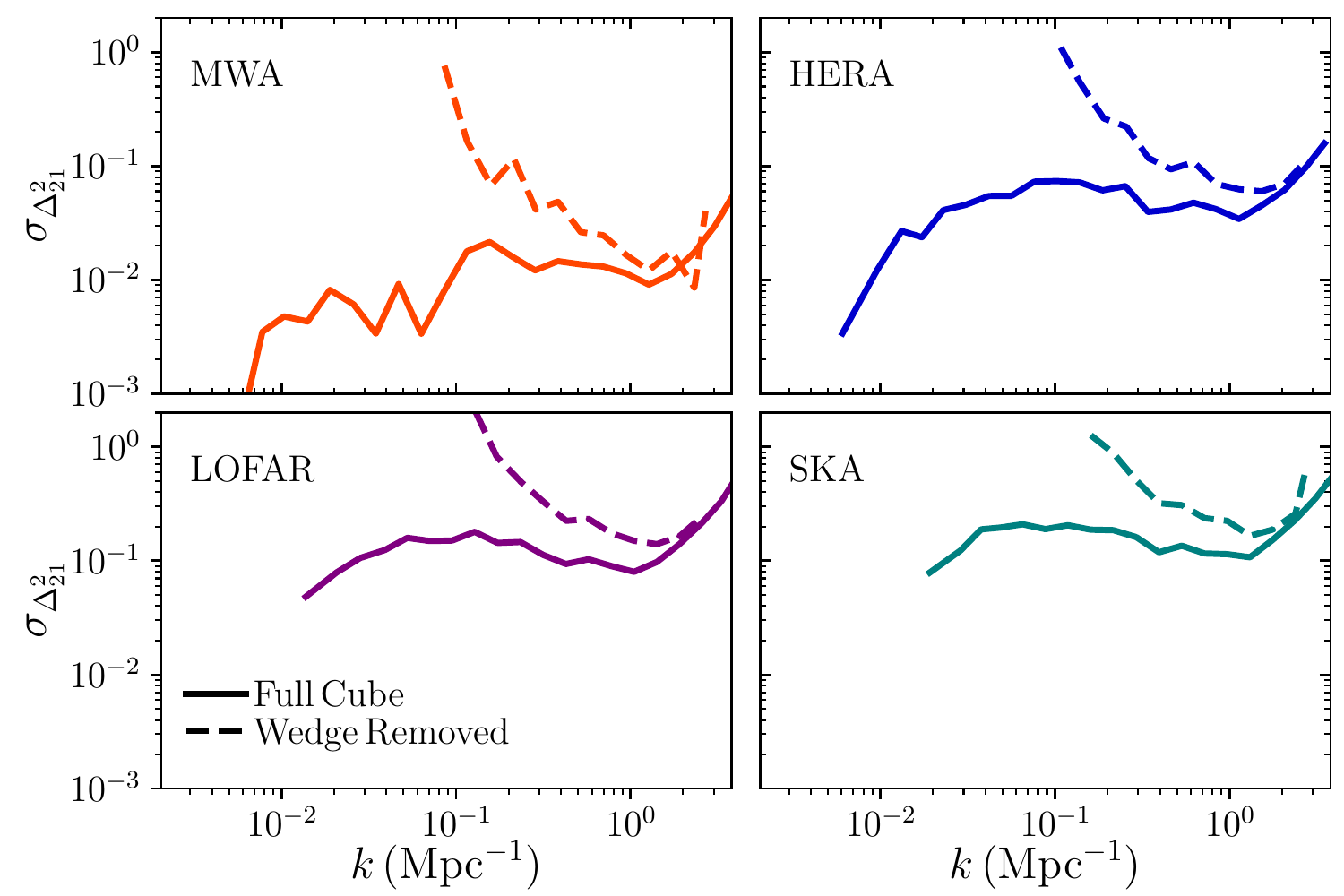}
	\end{center}
\caption[]{The $1\sigma$ variation in the 21-cm power spectrum estimated from sub-volumes of size equivalent to the 21-cm experiment field-of-view when using the full simulation box (solid curves) and after removing the foreground contaminated `wedge' modes (dashed). \textit{Top left:} MWA, \textit{Top right:} HERA, \textit{Bottom left:} LOFAR and \textit{Bottom right:} SKA.}
\label{fig:VariationWedge}
\end{figure*}

In Figure~\ref{fig:VariationWedge} we compare the 1$\sigma$ cosmic variance uncertainty after `wedge' avoidance (dashed curves) relative to the 21-cm PS estimates from the full simulation volume (solid curves). As expected, in avoiding the `wedge' region, we limit the total number of Fourier modes available in each spherical shell to measure the 21-cm PS. Thus, quite simply, our Poisson uncertainty increases owing to the smaller number of available modes driving the increasing uncertainty error. Further, as the number of available modes quickly diminishes as we move to larger spatial scales (smaller $k$) the cosmic variance uncertainty rapidly increases. For smaller scales, the cosmic variance uncertainty still becomes dominated by the Trispectrum component (as indicated by similar upticks in amplitude at $k>1$~Mpc$^{-1}$ as highlighted in Figure~\ref{fig:Variation}), though on slightly smaller scales than found previously. Nevertheless, the non-Gaussianity in the cosmic variance uncertainty is still important when performing `wedge' avoidance. 

Importantly, the differences between the two cases are amplified by our survey geometry. In general, our simulation volumes are considerably larger along the transverse direction relative to the line-of-sight. Thus, previously, when measuring the 21-cm PS over the full volume, our number statistics for the $k_{\perp}$ modes were considerably larger than for the $k_{\parallel}$ modes. Now, when measuring the 21-cm PS using foreground avoidance, most of these $k_{\perp}$ modes are excluded since they fall within the `wedge'.

In following this avoidance strategy, we also significantly limit the largest scale modes accessible to our 21-cm PS measurement. In all cases, the 21-cm PS is only measurable down to $k\sim0.1$~Mpc$^{-1}$, with no measurable modes beyond this limit. This limit is simply set by the adopted size of our buffer ($b = 0.1 \,h$~Mpc$^{-1}$) from Equation~\ref{eq:wedge} which is somewhat arbitrary. In principle, it can be reduced by considering either larger bandwidths or with an improved understanding of our systematics. Note however, that the smallest accessible scale for each instrument is not exactly $k\sim0.1$~Mpc$^{-1}$, but rather slightly larger. As we additionally ignore all $k_{\perp}=0$ modes, from Equation~\ref{eq:wedge} the minimum scale is simply $b$ plus $m$ times the smallest non-zero $k_{\perp}$ in our simulation volume. Thus, the minimum $k$ for the MWA is smaller than the SKA due to the smaller $k_{\perp}$ (larger transverse scale).

\begin{figure} 
	\begin{center}
	  \includegraphics[trim = 0.4cm 0.5cm 0cm 0.5cm, scale = 0.55]{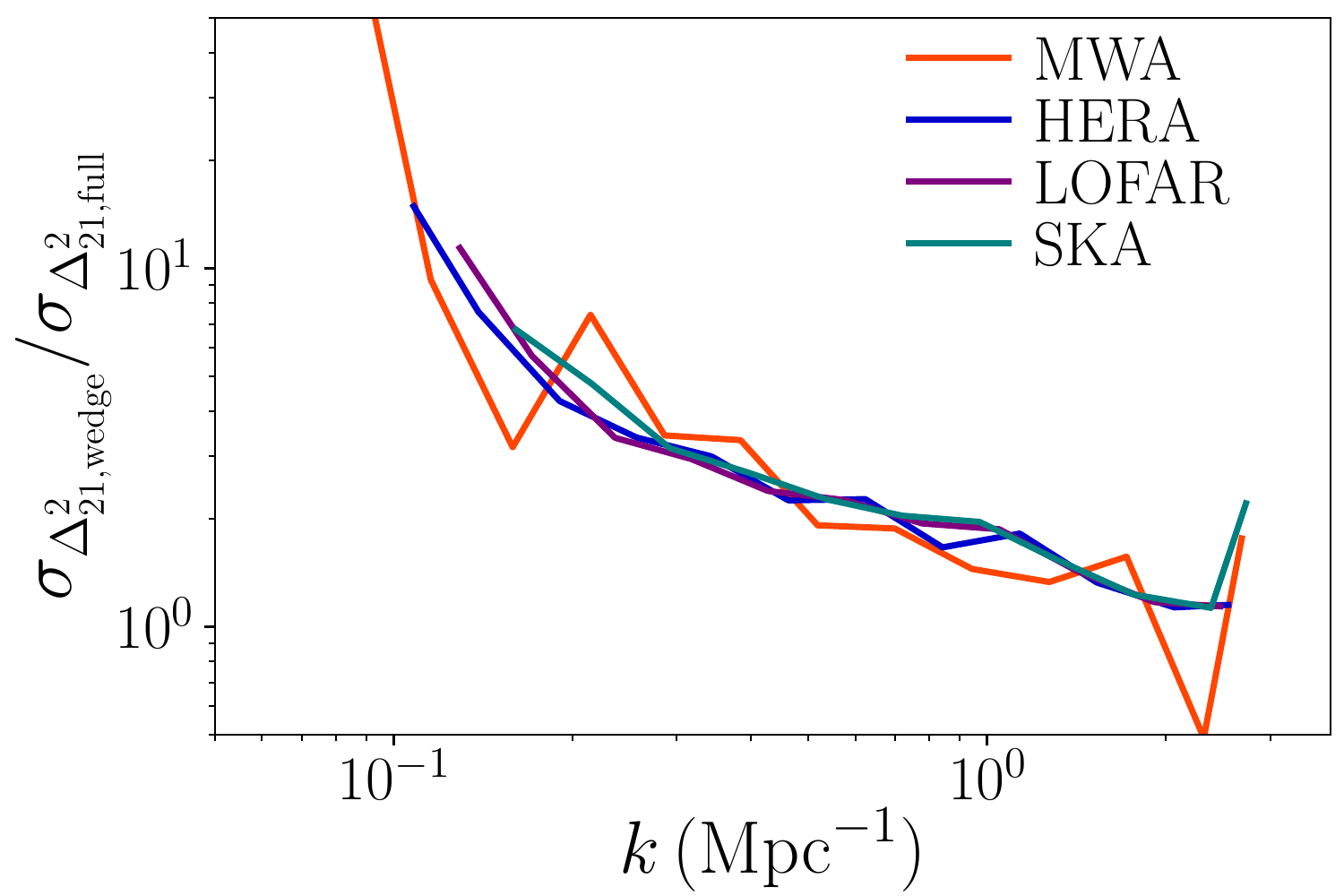}
	\end{center}
\caption[]{The fractional increase in the cosmic variance on the 21-cm power spectrum when removing the foreground wedge contaminated modes ($\sigma_{\Delta^{2}_{21,{\rm wedge}}}$) compared to the cosmic variance estimated from the full 21-cm simulation ($\sigma_{\Delta^{2}_{21,{\rm full}}}$).}
\label{fig:VariationWedge_Frac}
\end{figure}

In Figure~\ref{fig:VariationWedge_Frac} we present the fractional increase in cosmic variance when removing the foreground contaminated wedge modes. We find that the relative increase is comparable across all scales irrespective of the total simulation volume, highlighting that the increase in the cosmic variance is primarily driven by the fraction of excised Fourier modes (i.e. increased Poisson noise in each bin due to having fewer modes free of foregrounds). 

\subsection{Biased estimates of the 21-cm PS following `wedge' avoidance} \label{sec:wedge}

Previously, we explored the impact of foreground avoidance on the resultant cosmic variance uncertainty on a 21-cm PS measurement. Next, we consider whether the 21-cm PS measured following foreground avoidance can be biased relative to the 21-cm PS measured from a full simulation volume. This is particularly pertinent given that recent upper limits on the 21-cm PS measured by LOFAR \citep{Mertens:2020}, MWA \citep{Trott:2020} and HERA \citep{HERA:2022a} have for the first time allowed Bayesian astrophysical parameter inference to be employed to understand the physics of reionisation \citep{Ghara:2020,Ghara:2021,Greig:2021MWA,Greig:2021LOFAR,Mondal:2020,HERA:2022b}. However, when these inference pipelines are applied to these upper-limits, the simulated 21-cm PS is generated from the full simulation volume, rather than the characteristics of the observation, such as foreground avoidance. 

While likely irrelevant for the present upper-limits since these are not very aggressive and only disfavour relatively extreme models of reionisation. Nevertheless, as we approach a first detection, these inference pipelines will be vital to building our understanding of the galactic physics driving reionisation. Thus, it is important to explore the relative impact of any biases on the modelled 21-cm PS when not accounting for the characteristics of the observation.

Firstly, the observed 21-cm PS is asymmetric owing to redshift-space distortions from the peculiar motions of the gas along the line-of-sight. Thus unlike in the earlier sections, we must include the effects of redshift-space distortions on our estimate of the 21-cm brightness temperature signal (note, the inference pipelines above include these effects). These redshift-space distortions increase the relative power along the line-of-sight direction, which could amplify any biases when measuring the 21-cm PS following foreground avoidance. This is because modes are preferentially excised transverse to the line-of-sight, resulting in a higher amplitude signal after binning over all possible modes within spherical shells. Thus here, we compute the 21-cm brightness temperature signal for the full simulation following the multiplication of Equation~\ref{eq:Tb} by the pre-factor $\left(1 + \frac{{\rm d}v_{\rm r}}{H{\rm d}r}\right)^{-1}$ and computing the anisotropic 21-cm PS.

\begin{figure*} 
	\begin{center}
	  \includegraphics[trim = 0.8cm 0.6cm 0cm 0.3cm, scale = 1.]{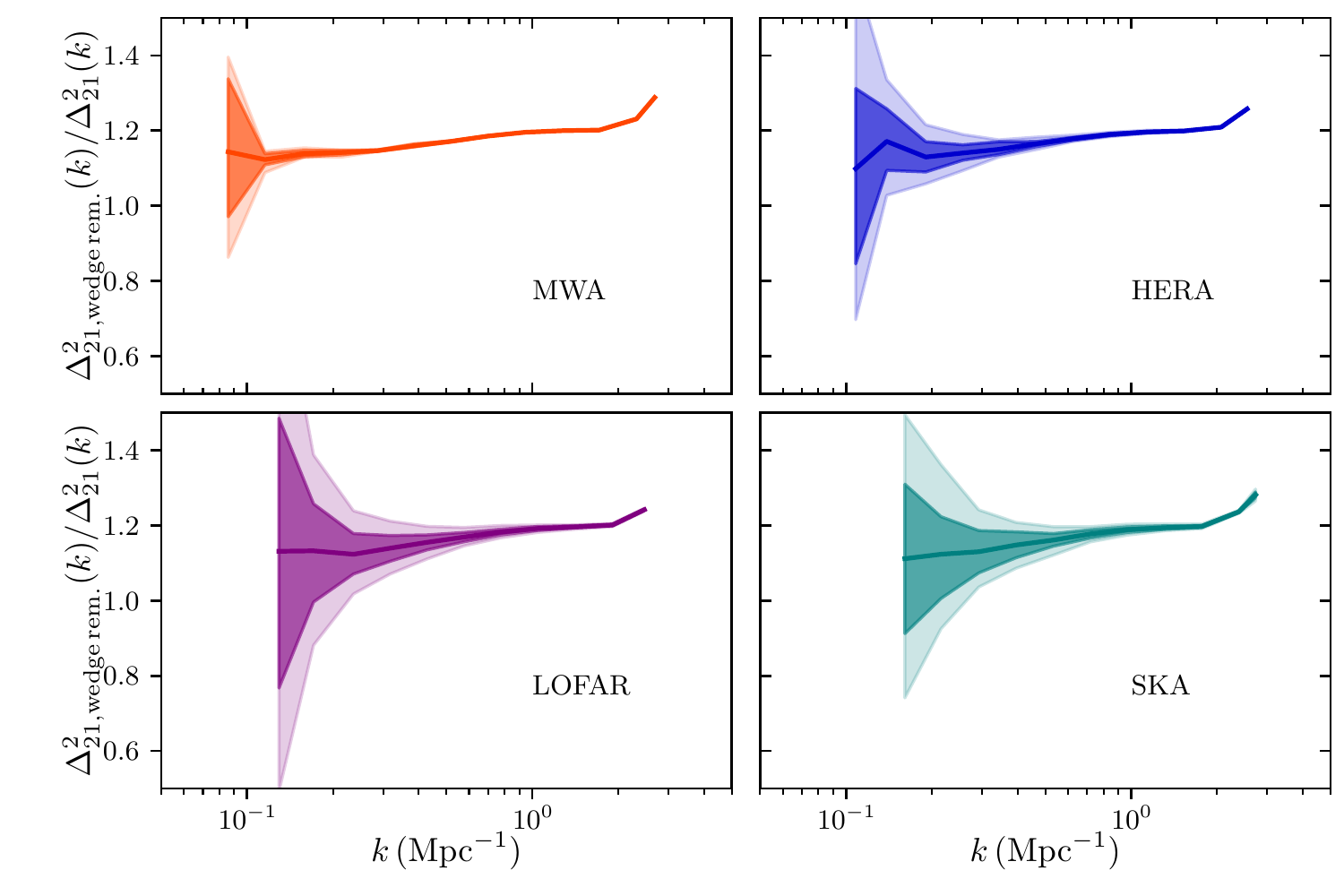}
	\end{center}
\caption[]{The ratio of the 21-cm PS with wedge modes removed (foreground avoidance) relative to the 21-cm PS from the full simulation volume. Solid curves correspond to the mean ratio, while dark and light shaded regions correspond to the 68th and 95th percentiles respectively. \textit{Top left:} MWA, \textit{Top right:} HERA, \textit{Bottom left:} LOFAR and \textit{Bottom right:} SKA.}
\label{fig:Ratio}
\end{figure*}

In Figure~\ref{fig:Ratio} we present the mean (solid curve) and the 68th and 95th percentiles (dark and light, respectively) for the ratio of the 21-cm PS with the `wedge' removed (avoidance) relative to the 21-cm PS from the full simulation volume. Here, these are determined using the same sub-volumes as obtained previously for each individual experiment from the full 7.5 Gpc simulation. It is clearly evident that for all survey areas, the 21-cm PS measured following foreground avoidance returns a biased estimate of the true 21-cm PS obtained from the full simulation volume. On average, we find the relative amplitude of this bias to be relatively modest, at $\sim10-20$ per cent over all measured $k$-scales. On the largest scales (small $k$), when foreground avoidance has limited sampling of the Fourier modes (see Figure~\ref{fig:VariationWedge}) this can result in increases/decreases in the 21-cm PS amplitude of larger than $40$ per cent.

Although fairly modest in amplitude, it is important to note that these results are model dependent. That is, this bias corresponds specifically to our particular source parameterisation and frequency band ($z\sim7.5-9.4$). Therefore, it is plausible that the amplitude of this bias could increase under different astrophysical models or stages of reionisation. Hence, this could have important consequences for parameter inference studies if the statistics of the simulated models are not consistent with the characteristics of the observation, potentially biasing the inferred astrophysical parameters. Although the model dependency of the work is important to consider here, in previous sections it is less relevant. When investigating the cosmic variance uncertainty the relative amplitude of the uncertainty is dependent on the number of modes within each spherical shell when estimating the 21-cm PS, not the 21-cm PS amplitude itself.

\begin{figure} 
	\begin{center}
	  \includegraphics[trim = 0.8cm 0.6cm 0cm 0.5cm, scale = 0.57]{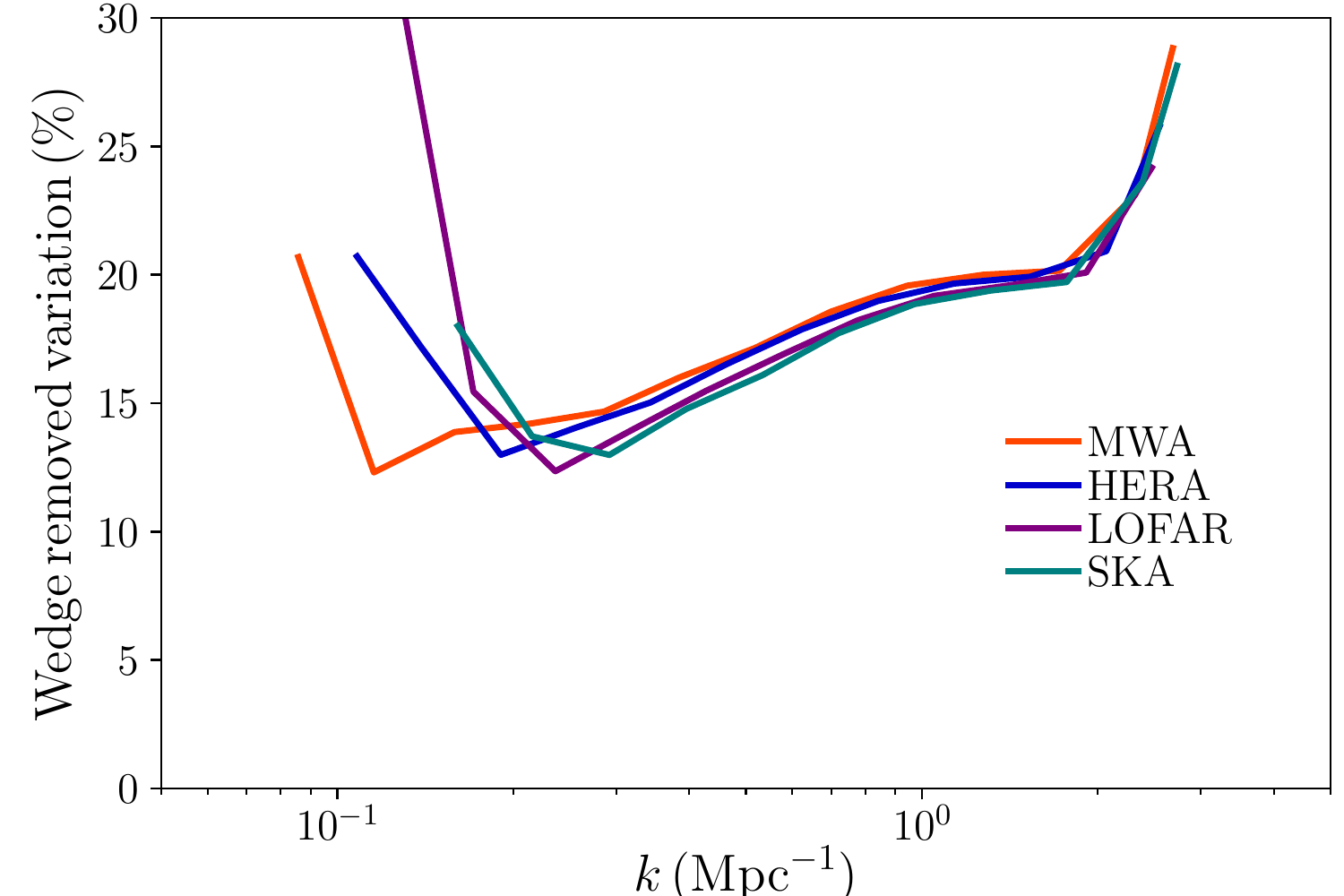}
	\end{center}
\caption[]{The mean ratio in the 21-cm PS with wedge modes removed compared to the 21-cm PS from the full simulation volume. The red, blue, purple and teal curves correspond to simulation boxes with fields of view equivalent to the MWA, HERA, LOFAR and the SKA respectively.}
\label{fig:Ratio_Combined}
\end{figure}

To more clearly illustrate the relative increase in amplitude in the 21-cm PS following foreground avoidance, in Figure~\ref{fig:Ratio_Combined} we show in the same panel the mean ratio of the 21-cm PS with/without foreground avoidance for all interferometer experiments. This clearly demonstrates that the relative increase in the 21-cm PS amplitude following foreground avoidance is consistent at $10-20$ per cent across all the different survey footprints. 

This $10-20$ per cent bias is broadly consistent with a similar investigation performed by \citet{Jensen:2016}. Here, this bias was determined for two astrophysical models over the full reionisation history. For their fiducial model, which more closely matches our model, at a similar stage of reionisation, $\bar{x_{\hi{}}}\sim0.5$, they find a similar bias of $10-20$ per cent, however, as a decrement rather than an amplification. Likely, these differences arise due to the specific modelling of the ionising sources, but more likely due to the simplifications present in our modelling (e.g. linear theory) relative to their $N$-body simulations. Importantly, though, they only consider a single realisation for this exploration. For sub-volumes equivalent in size to their single simulation (corresponding to the LOFAR field-of-view) in the lower left panel of Figure~\ref{fig:Ratio} we can accommodate individual models with a negative bias. Factoring in the modelling differences mentioned above, we are confident that these two works are broadly consistent.

Note, there also appears to be a slight gradient in the relative bias for the different experiments. The SKA, with the smallest survey footprint has the lowest amplitude whereas the MWA, with the largest footprint has the highest amplitude. However, this is simply due to the geometry of the survey volume. Foreground avoidance with the MWA removes on average more Fourier modes in each spherical shell below the `wedge' than that of the SKA relative to the 21-cm PS computed from the full simulation volume. This is because the MWA has a considerably larger transverse size (higher sampling of transverse modes) which results in a higher fraction of modes cut at the same fixed $k$ than the SKA. Equally, the uptick at the largest scales occurs earliest for the SKA compared to the other, larger area surveys. This occurs for the same reason as discussed earlier, the removal of the $k_{\perp}=0$ modes coupled with the $k_{\parallel}$ needing to satisfy Equation~\ref{eq:wedge}.

\subsection{Impact of tiling reionisation simulations}

Typically, when it is infeasible to generate a single, large volume simulation, the requisite size and volume can be achieved by stitching together many smaller simulations generated using periodic boxes \citep[see e.g.][]{Nasirudin:2020}. Although sufficient volumes are achieved in this manner, the statistics of the simulations will only be accurate on the scales modelled by the smaller simulation. That is, it is impossible to inject information on scales beyond the largest scale of the smallest simulation. Using our 7.5 Gpc simulation, we can explore the relative impact of tiling simulations on the large-scale information in the 21-cm PS.

The tiling of periodic simulations additionally leads to repeating structures on scales characteristic of the simulation volume. When considering discrete objects such as galaxies in a redshift survey, these repeating structures can be minimised through rotating the smaller simulations prior to tiling. Unfortunately, since the 21-cm signal arises from the neutral gas, it is a continuous rather than discrete quantity preventing the usage of rotations. Rotations would lead to edge effects and disjoint ionised regions on the boundaries of the smaller simulations leading to strange artefacts in the statistics. Thus, for 21-cm studies, we can only consider tiled simulations.

To investigate the impact on the large-scale power due to tiling, we consider three smaller volumes and tile each to match our single 7.5 Gpc, 6400 voxel simulation volume. We generate 50 independent realisations of the smaller simulations with 200, 400 and 600 voxels per side length, corresponding to lengths of $\sim234.4, 468.8$ and 703.1 Mpc. We tile the same individual simulation to the large volume footprint, thus we obtain 50 independent large volume tiled simulations. For each, we generate the full 21-cm light-cone and only consider a line-of-sight depth of 30 MHz, consistent with large volume simulation.

\begin{figure*} 
	\begin{center}
	  \includegraphics[trim = 2.5cm 0.6cm 0cm 0.5cm, scale = 0.54]{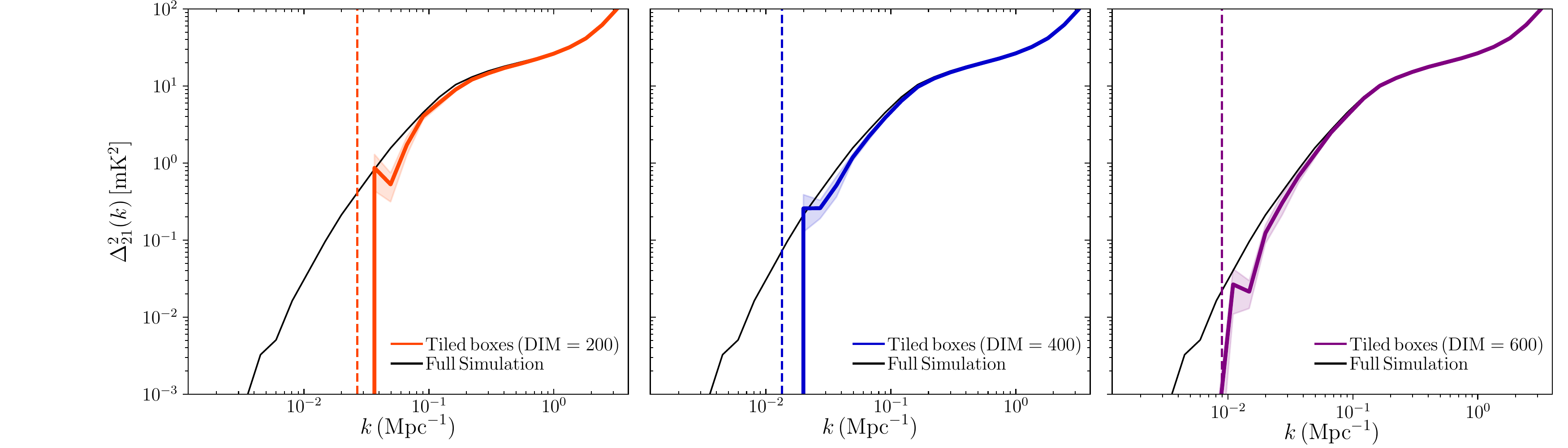}
	\end{center}
\caption[]{The 21-cm power spectrum obtained after tiling smaller volume simulations. The black curve corresponds to the 21-cm power spectrum of the full $7500 \times 7500 \times 500$ Mpc simulation, whereas the red, blue and purple curves correspond to the 21-cm power spectrum obtained from an equivalent simulation volume after tiling a single simulation box containing 200, 400 and 600 voxels per side length, respectively. Vertical dashed lines correspond to the fundamental frequency, $2\pi/L_{\rm small}$, for each small volume.}
\label{fig:Tiling}
\end{figure*}

In Figure~\ref{fig:Tiling} we compare the 21-cm PS obtained from our 7.5 Gpc simulation (black curve) to the 21-cm PS obtained from an equivalent tiled 7.5 Gpc simulation using the 200 (red), 400 (blue) and 600 (purple) voxel smaller simulations. In each, the vertical dashed line corresponds to the largest scale (fundamental frequency) sampled by the smaller simulations ($k_{f} = 2\pi/L_{\rm small}$). The shaded region corresponds to the 68th percentile scatter obtained from our 50 independent realisations. As expected, the 21-cm power drops away on scales below the fundamental frequency. This highlights that information on modes not originally simulated by the smaller volume simulations cannot be artificially created to match the full large volume simulation. Thus, a tiled simulation volume has a limited regime of validity restricted by the scale of the smaller simulations. 

On scales just above the fundamental frequency, we find the 21-cm power in the tiled box underestimates the power from the full simulation. In some instances this can be as much as a factor of $\sim2-3$. Although the physical scale of these spherically averaged Fourier modes are sampled by the original smaller volume simulation, on these scales we are still missing the contribution from Fourier modes which contain at least one component mode longer than the small volume simulation (e.g. $k_{x}, k_{y}$ or $k_{z} < k_{f}$). These individual modes contain no power, but contribute to the total number of modes within the bin, resulting in an overall reduction to the averaged power in that bin. Thus, the 21-cm power from the tiled box will underestimate the full box power on all scales until the fraction of Fourier modes with individual components larger than the smaller box is sufficiently small. Therefore, care must be taken to adequately produce sufficiently sized smaller volumes prior to tiling to minimise the loss of power on the largest scales.

In addition to underestimating the 21-cm power following the tiling of smaller periodic simulations, it is also important to note that we will underestimate the cosmic variance uncertainty in our tiled simulation volume. Since in the act of tiling we produce repeated copies of the spatial information from our smaller simulations we cannot artificially increase the variance within a sampled Fourier bin. That is, the measured sample variance is fixed to the variance (i.e. Poisson fluctuations) measured from the smaller volume simulation. However, since we have scaled up our total volume, following from Equation~\ref{eq:CVwT} the sample variance decreases as $\propto 1/V$. Thus our estimated cosmic variance from our tiled, large-volume simulation must be scaled by the increased volume. In effect, the tiled simulation underestimates the cosmic variance error by $\propto\sqrt{N_{\perp}^2\times N_{\parallel}}$, where $N_{\perp}$ is the number of copies required to achieve the desired transverse scale and $N_{\parallel}$ is the number of copies required along the line-of-sight. 

\begin{figure*} 
	\begin{center}
	  \includegraphics[trim = 2.5cm 0.6cm 0cm 0.5cm, scale = 0.54]{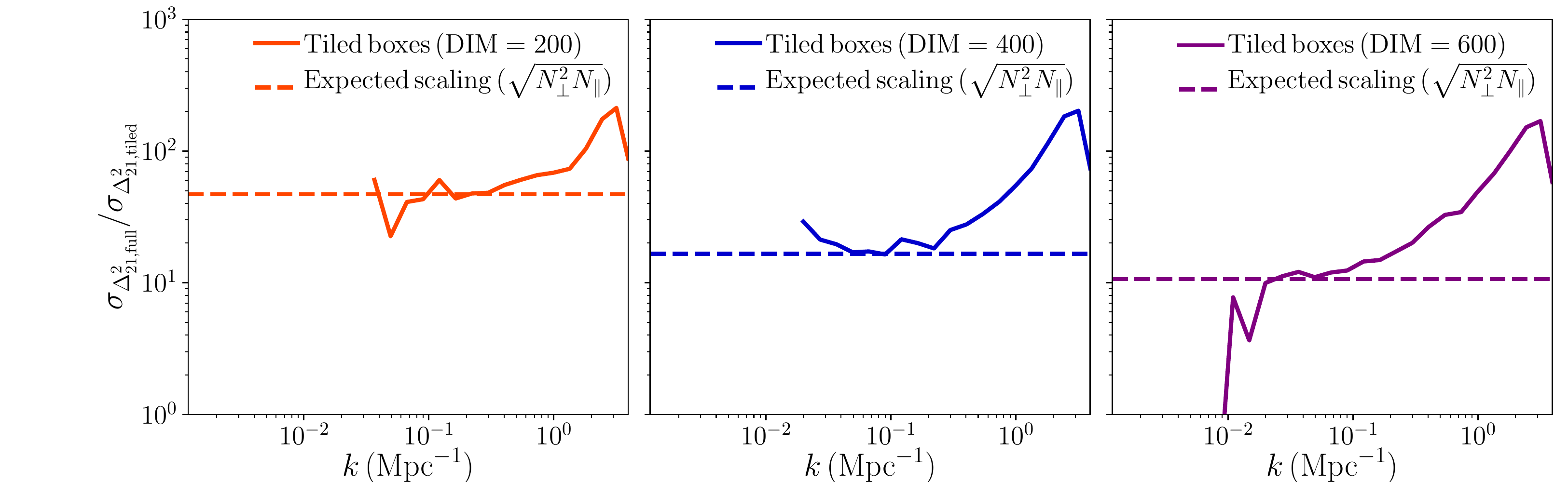}
	\end{center}
\caption[]{The ratio of the cosmic variance estimated from the full $7500 \times 7500 \times 500$ Mpc simulation to the cosmic variance estimated from tiling smaller volume simulations to achieve comparable volumes. Again, we consider  tiling a single simulation box containing 200 (red), 400 (blue) and 600 (purple) voxels per side length. The cosmic variance of the tiled simulations underestimates the true cosmic variance by an amount proportional to the total number of copies required to be tiled ($\propto\sqrt{N_{\perp}^2\times N_{\parallel}}$) as estimated by the horizontal dashed curves (see text for further details).}
\label{fig:TilingCV}
\end{figure*}

In Figure~\ref{fig:TilingCV} we highlight this scaling by showing the ratio of the cosmic variance uncertainty from the full simulated volume compared to that obtained from tiling the smaller volume simulations to achieve the same volume. In each, the tiled simulation cosmic variance underestimates the true cosmic variance, as indicated by the ratio sitting well above unity. The dashed horizontal lines correspond to the expected scaling ($\propto\sqrt{N_{\perp}^2\times N_{\parallel}}$) by which we underestimate the true cosmic variance in each tiled simulation volume. That is, the cosmic variance in the tiled simulation must be decreased by this amount to match the true cosmic variance from the full simulation volume. For larger $k$ (i.e. $k>0.5$~Mpc$^{-1}$) the impact of the non-Gaussian (Trispectrum) contribution to the cosmic variance becomes increasingly important, modifying this simple scaling relation.

Importantly, here we have only considered the power extracted from the full simulation volume. However, if one also considers the instrumental response of the beam (i.e. beam multiplied by the tiled signal), then additional artefacts in the estimated power could be expected due to mode mixing and/or power leaking in from the side-lobes. We leave an investigation into these effects to future work.

\section{Conclusion} \label{sec:Conclusion}

In preparation for the wealth of data expected from large-scale interferometer experiments such as the MWA, LOFAR, HERA and the SKA, we must rigorously test and validate the full data analysis and reduction pipelines. For experiments such as the MWA, with a field-of-view of $\sim25^{2}$ deg$.^{2}$ at 150 MHz ($z\sim8.5$) this requires reionisation simulations in excess of $\sim4$~Gpc. To explore the effects of side-lobes away from the primary beam and the potential for large-scale power leakage, even larger volumes are required. 

We introduced a modified version of the semi-numerical reionisation simulation code \cmfst{} \citep{Mesinger:2007p122,Mesinger:2011p1123} specifically tailored towards generating these extremely large volume simulations. To achieve this, we limit the line-of-sight depth of the simulations to specific observing bandwidths (though fully sample the large-scale modes), allowing the transverse scales to be extremely large. Further, we forego some level of accuracy by limiting structure formation to linear theory. However, for the purposes of testing and validation, provided the simulated 21-cm signal behaves in a similar manner (i.e. correct statistics and correlated behaviour) to the expected signal, these assumptions are sufficient. Using this, we generate a 30 MHz band-width ($\sim500$~Mpc) simulation with 6400 voxels over a 7.5 Gpc transverse side-length ($\sim1.17$ cMpc resolution). This simulation was specifically tailored for the recent 2020 MWA upper-limits \citep{Trott:2020}

With such a large volume simulation in hand, we then explored the statistical behaviour of the 21-cm power spectrum (PS) and assumptions made in the literature when computational limitations restrict the maximum achievable simulation volumes. We performed these using simulation volumes equivalent in size to the field-of-view for each of the 21-cm interferometer experiments, MWA, LOFAR, HERA and the SKA. We found:
\begin{itemize}
\item finite simulation volumes do not exhibit any loss or increase in large-scale power from missing modes longer than the finite simulation volume.
\item the cosmic variance uncertainty on a 21-cm PS measurement at $k\gtrsim0.5$~Mpc$^{-1}$ is dominated by the non-Gaussianity in the 21-cm signal by $\gtrsim10\times$ over that expected under the typical Gaussian assumption. This is consistent with that found earlier by \citet{Mondal:2016}. 
\item the excision of contaminated foreground wedge modes significantly increases the amplitude of cosmic variance due to the reduction in the number of available Fourier modes. Further, it restricts the largest scale modes accessible.
\item measuring the 21-cm PS following wedge mode excision results in a modestly biased estimate (increase of $\sim10-20$ per cent) of the true underlying 21-cm PS. Though, the relative amplitude will be heavily model dependent. This bias is broadly consistent with that found by \citet{Jensen:2016} and should be an important consideration for reionisation astrophysical parameter inference studies.
\item tiling smaller simulations to achieve extremely large volumes results in an underestimate of the 21-cm power on scales smaller than the small box fundamental frequency $k_{f} \gtrsim 2\pi/L_{\rm small}$. Therefore, care must be taken when constructing smaller volume simulations for tiling to ensure the relevant scales of interest are sufficiently sampled.
\item tiling smaller simulations additionally results in underestimating the true cosmic variance of the desired large volume simulation. The underestimation can be compensated for by scaling the uncertainty by $\propto\sqrt{N_{\perp}^2\times N_{\parallel}}$ which accounts for the number of repeated copies of the small volume simulation required to match the transverse and line-of-sight scales, respectively.
\end{itemize}

\section*{Acknowledgements}

We thank Garrelt Mellema, Rajesh Mondal and Andrei Mesinger for useful discussions relating to this work. Parts of this research were supported by the Australian Research Council Centre of Excellence for All Sky Astrophysics in 3 Dimensions (ASTRO 3D), through project number CE170100013. CMT is supported by an ARC Future Fellowship under grant FT180100321. Parts of this work were performed on the OzSTAR national facility at Swinburne University of Technology. OzSTAR is funded by Swinburne University of Technology.

\section*{Data Availability}

The data underlying this article will be shared on reasonable request to the corresponding author.

\bibliography{Papers}

\end{document}